\documentclass{article}

\usepackage{arxiv}

\usepackage[utf8]{inputenc} 
\usepackage[T1]{fontenc}    
\usepackage{hyperref}       
\usepackage{url}            
\usepackage{booktabs}       
\usepackage{amsfonts}       
\usepackage{amsmath}
\usepackage{nicefrac}       
\usepackage{microtype}      
\usepackage{graphicx}
\usepackage{multirow}

\newcommand{\comment}[1]{}

\newcommand{\degree}{^{\circ}}
\usepackage{doi}
\usepackage[
backend=biber,
style=nature,
]{biblatex}

\DeclareSourcemap{
  \maps[datatype=bibtex]{
    \map{
      \step[fieldset=issn, null]
      \step[fieldset=series, null]
      \step[fieldset=note, null]
      \step[fieldset=language, null]
    }
  }
}

\defbibheading{bibliography}[\refname]{\paragraph{#1:}}

\author{ \href{https://orcid.org/0000-0002-0783-2489}{\includegraphics[scale=0.06]{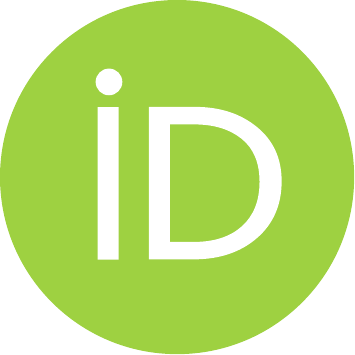}\hspace{1mm}Simon C. St\"ahler} \\
    Institute for Geophysics\\
	ETH Zürich\\
	\And
    \href{https://orcid.org/0000-0002-5603-7334 }{\includegraphics[scale=0.06]{orcid.pdf}\hspace{1mm}Anna Mittelholz} \\
	Department of Earth and Planetary Sciences\\
	Harvard University\\
    \And
	\href{https://orcid.org/0000-0002-7200-5682}{\includegraphics[scale=0.06]{orcid.pdf}\hspace{1mm}Cl\'ement Perrin} \\
	LPG-OSUNA \\
	Nantes Universit\'e\\
	
	\AND

    \href{https://orcid.org/0000-0001-5246-5561}{\includegraphics[scale=0.06]{orcid.pdf}\hspace{1mm}Taichi Kawamura} \\
	Université Paris Cit\'e\\
 Institut de physique du globe de Paris, CNRS \\
Paris, France 
	\And

     \href{https://orcid.org/0000-0003-4594-2336}{\includegraphics[scale=0.06]{orcid.pdf}\hspace{1mm}Doyeon Kim} \\
	Institute for Geophysics\\
	ETH Zürich\\
	\And
		\href{https://orcid.org/0000-0003-0319-2514}{\includegraphics[scale=0.06]{orcid.pdf}\hspace{1mm}Martin Knapmeyer} \\
	Institute for Planetary Science\\
	DLR Berlin\\
	\And

   \href{https://orcid.org/0000-0001-9401-4910}{\includegraphics[scale=0.06]{orcid.pdf}\hspace{1mm}G\'eraldine Zenh\"ausern} \\
	Institute for Geophysics\\
	ETH Zürich\\
	\And
	\href{https://orcid.org/0000-0001-8626-2703}{\includegraphics[scale=0.06]{orcid.pdf}\hspace{1mm}John Clinton} \\
	Institute for Geophysics\\
	ETH Zürich\\
	\And
	    \href{https://orcid.org/0000-0002-5573-7638}{\includegraphics[scale=0.06]{orcid.pdf}\hspace{1mm}Domenico Giardini} \\
	Institute for Geophysics\\
	ETH Zürich\\
	\And
    \href{https://orcid.org/0000-0002-1014-920X}{\includegraphics[scale=0.06]{orcid.pdf}\hspace{1mm}Philippe Lognonn\'e} \\
	Université Paris Cit\'e\\
 Institut de physique du globe de Paris, CNRS 	\And
    \href{https://orcid.org/0000-0003-3125-1542}{\includegraphics[scale=0.06]{orcid.pdf}\hspace{1mm}W. Bruce Banerdt} \\
	Jet Propulsion Laboratory\\
	California Institute of Technology

}

\renewcommand{\headeright}{S. C. St\"ahler et al.}
\renewcommand{\shorttitle}{Seismicity of Cerberus Fossae}

\hypersetup{
pdftitle={Marsquakes indicate dike-induced tectonics in Cerberus Fossae, Mars},
pdfsubject={Mars},
pdfauthor={Simon St\"ahler et al.},
pdfkeywords={Mars, Tectonics, Marsquakes},
}
\newcommand{\citep}{\cite}
\title{Marsquakes indicate dike-induced tectonics in Cerberus Fossae, Mars}

\begin{document}
\maketitle


\begin{abstract}
 On Mars, seismic measurements of the InSight lander have for the first time confirmed tectonic activity in an extraterrestrial geological system: the large graben system Cerberus Fossae. In-depth analysis of available marsquakes thus allows unprecedented characterization of an active extensional structure on Mars. We show that both major families of marsquakes, characterized by low and high frequency content, LF and HF events respectively, can be located on central and eastern parts of the graben system. This is in agreement with the decrease in structural maturity towards the East as inferred from orbital images. LF marsquake hypocenters are located at about 15-50 km and the spectral character suggests a weak, potentially warm source region consistent with recent volcanic activity at those depths. The HF marsquakes occur in the brittle shallow part and might originate in the fault planes associated with the graben flanks. One particularly intriguing cluster of seismic activity is located close to a recently identified active young volcano near Zunil crater. The inferred mechanically weak source region supports the existence of an active plume below Elysium Mons, from which dikes are extending into Elysium Planitia. We find no trace of seismic activity on compressional thrust faults on Mars, in opposition to the models of seismicity driven by secular cooling and lithospheric contraction.
\end{abstract}

Faults litter the martian surface \citep{knapmeyer_working_2006, tanaka_geologic_2014}, providing evidence for brittle deformation throughout the planet's history. Due to the lack of recent widespread volcanism, plate tectonics or high erosion rates which recycle the surfaces of Venus or Earth, they are well preserved over billions of years and do not necessarily correlate with recent tectonic deformation. The Interior Exploration using Seismic Investigations, Geodesy and Heat Transport mission (InSight) mission landed on Mars to (amongst other objectives) characterize current day seismicity and thus tectonic activity using a broadband seismometer \citep{lognonne_seis_2019, banerdt_initial_2020}. 
Around InSight's landing site, wrinkle ridges and lobate scarps, interpreted as buried reverse faults resulting from compression, are widely spread, with clusters in the large Isidis and Hellas impact basins, but also in the planes of Hesperia, Arcadia and Amazonis (Figure \ref{fig:overview_map}). Their abundance was interpreted as the result of secular cooling and associated shrinking of the planet \citep{phillips_expected_1991}, in combination with the weight of the large Tharsis volcanic region 6000 km to the East.  
Young (<600 Ma) normal (extensional) faults extend radially outwards from Tharsis. Westward, these are Cerberus Fossae \citep{berman_recent_2002} in Eastern Elysium Planitia, and further southward these are Memnonia and Sirenum Fossae \citep{anderson_primary_2001}. Therefore, based on the distribution of young faults one might expect widespread seismic activity north and south-east of the InSight landing site due to compression of the crust and extensional faulting associated with the three fossae (Figure \ref{fig:overview_map}). 
However, the first seismic data revealed a different picture.

The most significant marsquakes during InSight's first Martian year of operations (observed on Sols 173 and 235 of the mission, thus named \textit{S0173a} and \textit{S0235b}) were located at the approximate distance and in direction of Cerberus Fossae \citep{lognonne_constraints_2020, giardini_seismicity_2020}. Ever since, no other tectonic feature on the InSight hemisphere of Mars has been unequivocally confirmed to be seismically active \citep{clinton_marsquake_2021}, and only recently, on Sols 976 and 1000, InSight detected two large marsquakes on the far side, in the Tharsis province. No marsquakes at all have been located on wrinkle ridges or lobate scarps, i.e. compressional features so far.

Further analysis of seismic waveforms showed that respective source mechanisms of the large events from Sols 173 and 235 are consistent with an extensional setting \citep{brinkman_first_2021}, suggestive of ongoing opening of the Cerberus Fossae. Until Sol 1100 (2021/12/31), 18 out of 24 low-frequency (LF) marsquakes for which a location could be determined, have been located at a distance consistent with Cerberus Fossae \citep{clinton_marsquake_2021, insight_marsquake_service_mars_2021, zenhausern_lowfrequency_2022} and source mechanisms have been inverted for nine events in total and are compatible with an extensional setting \citep{jacob_seismic_2022}. LF quakes are similar in character to earthquakes, with clear P- and S-waves, and they are thought to occur in the lower crust or uppermost mantle \citep{giardini_seismicity_2020}, between 15 and 50 km depth \citep{brinkman_first_2021, stahler_seismic_2021}. A second class of marsquakes is termed high-frequency (HF) events, due to significant signal energy above 2 Hz \citep{clinton_marsquake_2021}. A long signal duration is interpreted as the result of a shallow hypocenter of less than a few km depth, which excites reverberations of seismic waves in shallow subsurface layers \citep{van_driel_high-frequency_2021, menina_energy_2021, karakostas_scattering_2021}. HF events have been detected in much larger numbers, 1150 until 2021/12/31, yet they have smaller magnitudes than LF events; their distance is clustered around 1500 km, and due to a lack of clear polarisation, their direction as seen from the lander has not been determined and thus no tectonic explanation has been provided so far. 

Cerberus Fossae was previously identified as the location with most recent volcanic activity on Mars, based on global surface age mapping \citep{tanaka_geologic_2014} and local analysis of young basalt deposits \citep{berman_recent_2002}. Those are dated to less than 10 Ma ago \citep{taylor_estimates_2013}, contemporary with the deposition of surficial flow units over Eastern Elysium Planitia \citep{voigt_investigating_2018}. Moreover, \cite{horvath_evidence_2021} identified symmetric dust deposits around some fossae that are suggestive of explosive volcanism within the last 200 ka.
In depth-analysis of available seismic data allows for the first time to corroborate the active nature of the Cerberus graben system. We present a first consistent interpretation of a tectonic system on another planet using seismic data, with implications for geodynamics of Mars and terrestrial planets in general.
\section*{Geological context}
The Cerberus Fossae graben system is approximately 1200-2300 km (20-40\textdegree) east of the InSight lander (Figure \ref{fig:CF_map}b,c). It consists of five main graben features (G1-G5 in Figure \ref{fig:CF_map}b) trending NW to SE. The five grabens are generally between 250 and 600 km long, but are further segmented. Smallest segments that can be identified on the surface are 5-10 km long \cite{perrin_geometry_2022}. 
The grabens disect the Cerberus Plains Unit and are constrained to be younger than 10 Ma based on crater counts \cite{taylor_estimates_2013, perrin_geometry_2022}. Signs of erosion indicate that the westernmost fractures are oldest. Here, the grabens are more maturely developed and well-connected, as opposed to the hardly connected segments at the eastern fractures \cite{perrin_geometry_2022}. 
A long-lasting discussion about the origin of the Cerberus Fossae led to three main scenarios: (1) The weight of the Tharsis volcanic province leads to large-scale radially extensional and concentrically compressional faulting \citep{banerdt_stress_1992}. 
(2) The Tharsis highlands are dynamically supported by a plume or degree-1 mantle convection. The radial faults are caused by giant dikes extending radially from this plume \citep{ernst_giant_2001}. (3) While the radial extensional faults are due to a global stress field influenced by (1), the occurrence of Cerberus Fossae specifically is due to a dike system extending from Elysium Mons \citep{hauber_very_2011}.

All three scenarios have implications for the observed seismicity. In the graben scenario (1), seismicity would be equi-distributed over the system and mainly show extensional focal mechanisms. The dike scenarios (2 and 3), would imply more localized seismicity where source characteristics show indications of weakened, warm material. In case (2), the dike tip and seismicity would be focused in the Western part of Cerberus Fossae, in case (3) in the Eastern parts. We will argue that the latter is indeed the case. 

\section*{Marsquake Locations}

\subsection*{Distance}
Marsquakes are located by the Marsquake Service (MQS; \citep{clinton_marsquake_2018}). Their respective distances from the InSight lander are determined purely from seismic data, without taking prior tectonic information into account. For LF events, MQS uses the arrival time difference between P- and S-waves, the two strongest seismic body waves to compute the distance of the event from InSight. This travel time difference is compared to predicted travel times for a suite of inferred one-dimensional velocity-density structure models of Mars' interior \citep{clinton_marsquake_2021, khan_imaging_2021, stahler_seismic_2021}. The uncertainty in absolute distance  is a combination of uncertainty in picking the arrival times and the span of possible seismic velocities in the interior model suite \citep{bose_probabilistic_2016}. The uncertainty in seismic velocities from most recent interior models is about 5\% \citep{khan_imaging_2021}. If one separates the effect of pick uncertainty and velocity model uncertainty, the relative distances of all marsquakes can be determined with much higher precision than their absolute distances, allowing to identify a cluster of seismic activity. The absolute distance of this cluster can then be estimated using different types of interior velocity models (Figure \ref{fig:CF_map}a). 

We use P and S-wave pick times of the MQS catalog version 9 \cite{insight_marsquake_service_mars_2021} for marsquakes in distances within 3000 km ($\approx 50\degree$) of InSight and investigate the distance spread resulting from 2 end member velocity models and one median model from \citep{stahler_interior_2021}. The model with slowest/fastest velocities creates the set of \textit{nearby}/\textit{far} solutions. From these event distances, we determine a normalized seismicity density over distance (Figure \ref{fig:CF_map}a). The uncertainty in depth of the events is estimated to be on the order of 20~km, and is reflected in the distance distributions.  

For HF events, we use MQS S/P picks (termed Sg and Pg, due to crustal propagation \citep{clinton_marsquake_2021}) and according to MQS practice we assume that the onsets of the two observed phases propagate with velocities of $v_{\text{Pg}}$=4 km/s and $v_{\text{Pg}}/v_{\text{Sg}}=\sqrt{3}$ \citep{clinton_marsquake_2021}, consistent with velocities of the lower crust \citep{knapmeyer-endrun_thickness_2021}. The 286 HF events spread over a distance range from 1200-2500 km (20 to 40\textdegree), with a clear maximum between 1700 and 2000 km (27 and 32\textdegree; see Figure \ref{fig:CF_map}a), in the center of Cerberus Fossae (Figure \ref{fig:CF_map}a). 

\subsection*{Direction}
Because InSight comprises a single-station global network, the direction towards an LF event (termed back-azimuth) is estimated independently from the distance. MQS uses the linearity of P-wave motion \citep{bose_probabilistic_2016}, which only resulted in direction estimates for 8 events, due to the low SNR. A recently proposed more robust alternative is based on the eigenvalues of the spectral matrix \citep{schimmel_use_2003} for P- and S-waves \citep{zenhausern_lowfrequency_2022}. With this approach, we found that at least 14 events are located within 150 km north/south of Cerberus Fossae as seen from InSight (see table 1). By application of the method to well-located terrestrial data of similar signal-to-noise ratio, \cite{zenhausern_lowfrequency_2022} found that this is within the uncertainty of the method at a distance of 1500 km. Therefore, all these 14 LF marsquakes are compatible with locations in Cerberus Fossae (Table \ref{tab:marsquakes}).

The direction of HF events has so far been unknown, because the highly scattered first arrival has not shown an increased degree of polarization for any single event \citep{zenhausern_lowfrequency_2022, clinton_marsquake_2021}. Here, we make use of the large number of HF events observed so far. To investigate whether the epicenters of the HF events are in a similar location, we stack all HF event waveforms and compare horizontal seismogram power in the radial direction of central Cerberus Fossae (70\textdegree from North, radial) with that in the orthogonal direction (transverse). If the sources are indeed located in this direction, we expect the P-wave arrival to show higher energy in radial direction, at least in a short time window, in which ballistic waves dominate. Figure \ref{fig:HF_direction}a shows no clear effect, likely due to time shifts between the phase arrivals of individual events. We follow \citep{kim_improving_2021} and conduct a realignment of the Pg arrivals using a multichannel cross-correlation method \citep{vandecar_determination_1990}. After alignment, we select 32 events in a distance between 23\textdegree and 25\textdegree to minimize the variation in backazimuth within the event stack. We find that the ratio of radial to transverse energy is maximized for a back-azimuth of $78\degree\pm12\degree$, supporting the identification of Cerberus Fossae as source region of the HF events.

\subsection*{Locating Seismicity within the Cerberus Fossae}
\begin{figure}
    \centering
    \includegraphics[width=0.7\textwidth]{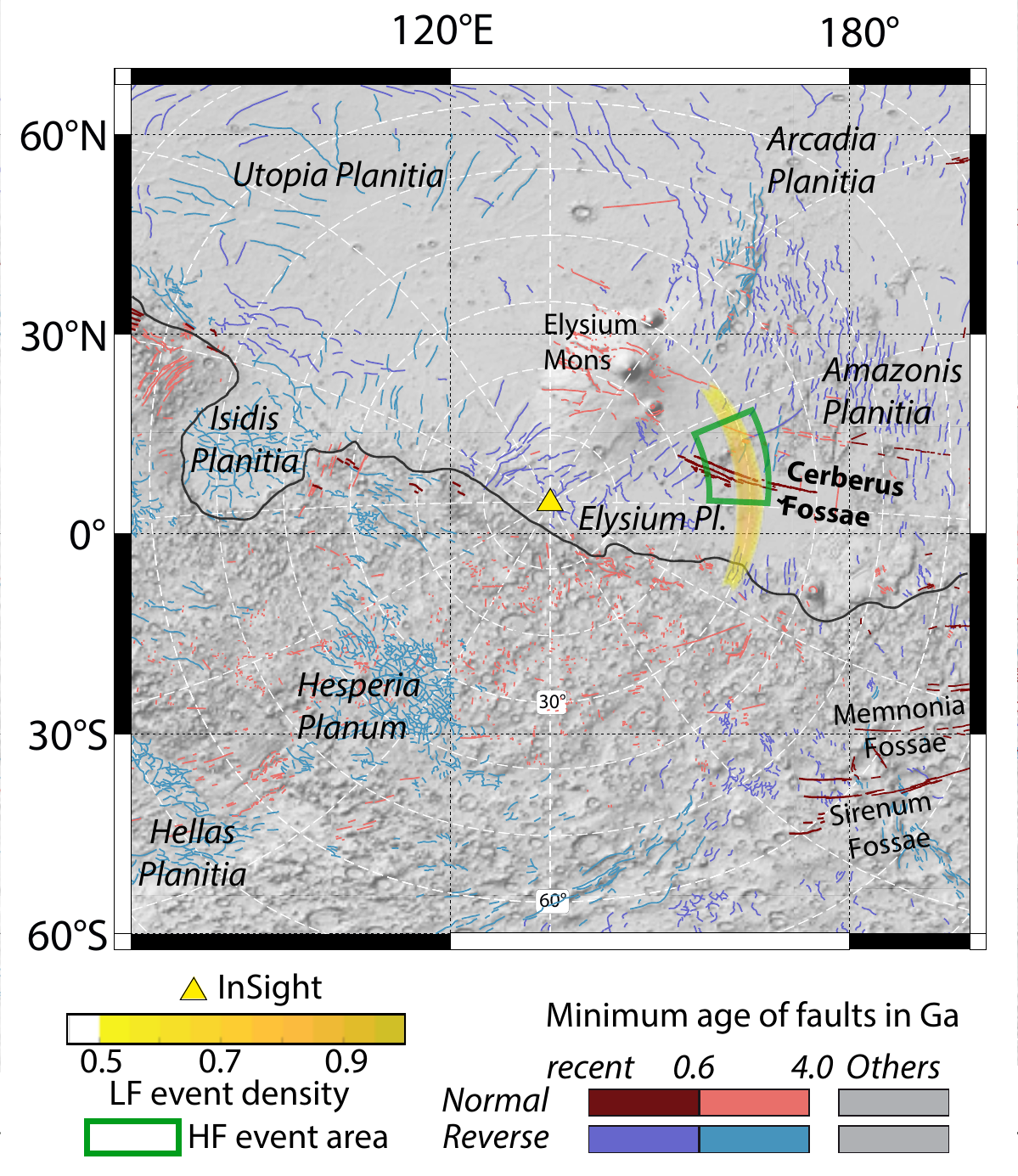}
    \caption{Faults and major geographic features around the InSight landing site \citep{knapmeyer_working_2006, perrin_geometry_2022} color-coded by age. The yellow shaded area marks the normalized density of low frequency (LF) quakes \citep{insight_marsquake_service_mars_2021, zenhausern_lowfrequency_2022}. The green box highlights the backazimuth range found by our analysis for HF marsquakes, corresponding to the dashed line in figure \ref{fig:HF_direction} and the distance range in which 80\% of seismicity is present.  
    Background shading: MOLA topographic map \citep{smith_mars_2001}. A global version of this map is available in the supplement.}
\label{fig:overview_map}
\end{figure}

\begin{figure*}
\centering
\includegraphics[width=.9\textwidth]{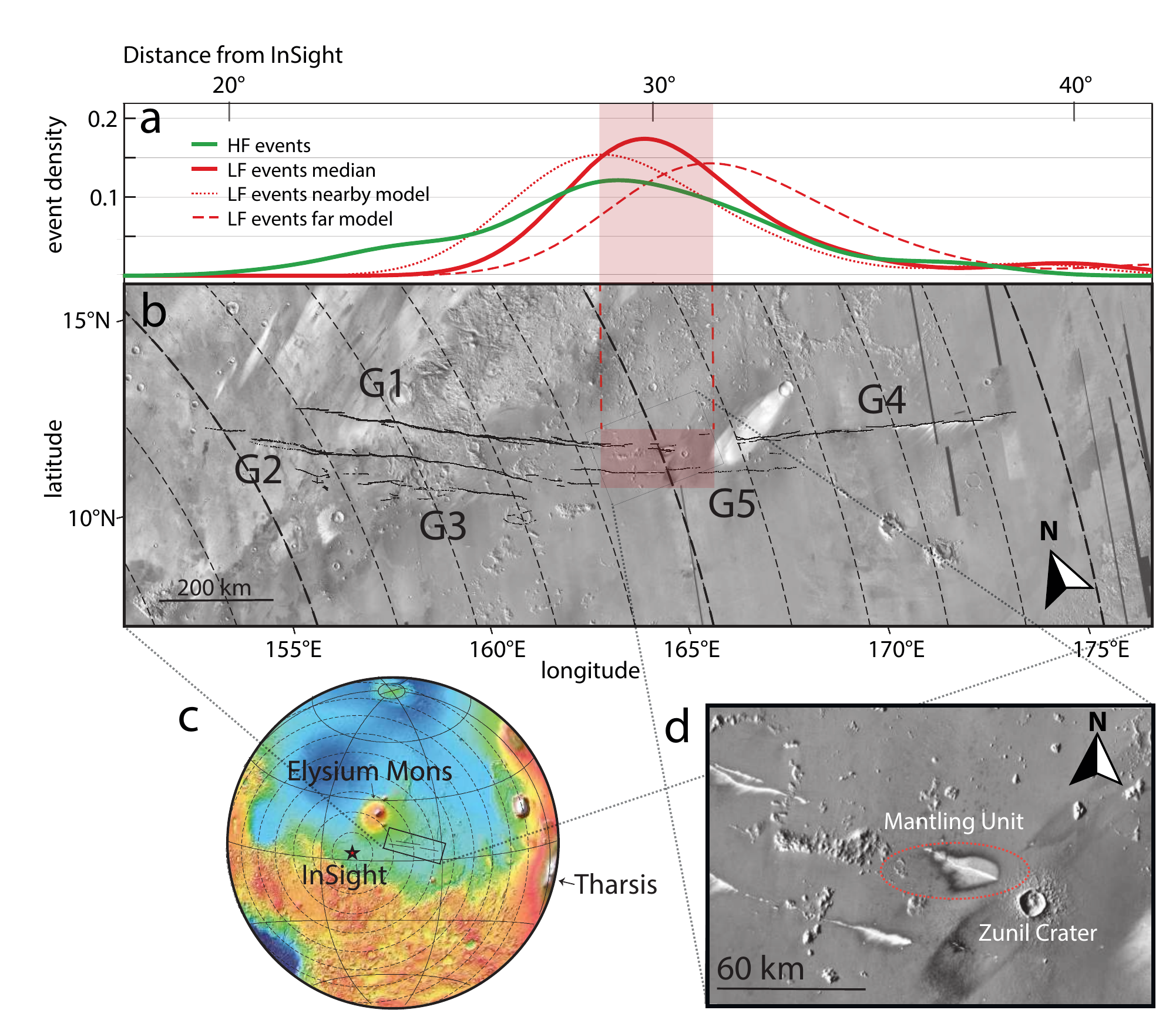}
\caption{(a) Density of seismic moment release along the profile of Cerberus Fossae. Probability density functions (PDF) from a Gaussian kernel density estimation of high frequency (HF; green) and low frequency (LF; red) moment are shown separately. For the LF events, 3 candidate distributions based on different mantle models are shown ("median" model is shown in Figure 1). The red box highlights the area between near and far model peaks in (a) for easier comparison with (b).  
(b) Oblique Mercator projection of Mars Odyssey's Thermal Emission Imaging System (THEMIS; day-time infrared) highlighting the 5 main graben features G1-G5 of Cerberus Fossae (mapping from \citep{perrin_geometry_2022}).
(c) MOLA topography inset for global context and 10 degree distance circles around InSight. (d) The area of highest marsquake density: The Cerberus Mantling unit recently identified as a volcanically active feature by \citep{horvath_evidence_2021} is circled.}
\label{fig:CF_map}
\end{figure*}

\begin{figure}
\centering
\includegraphics[width=0.75\linewidth]{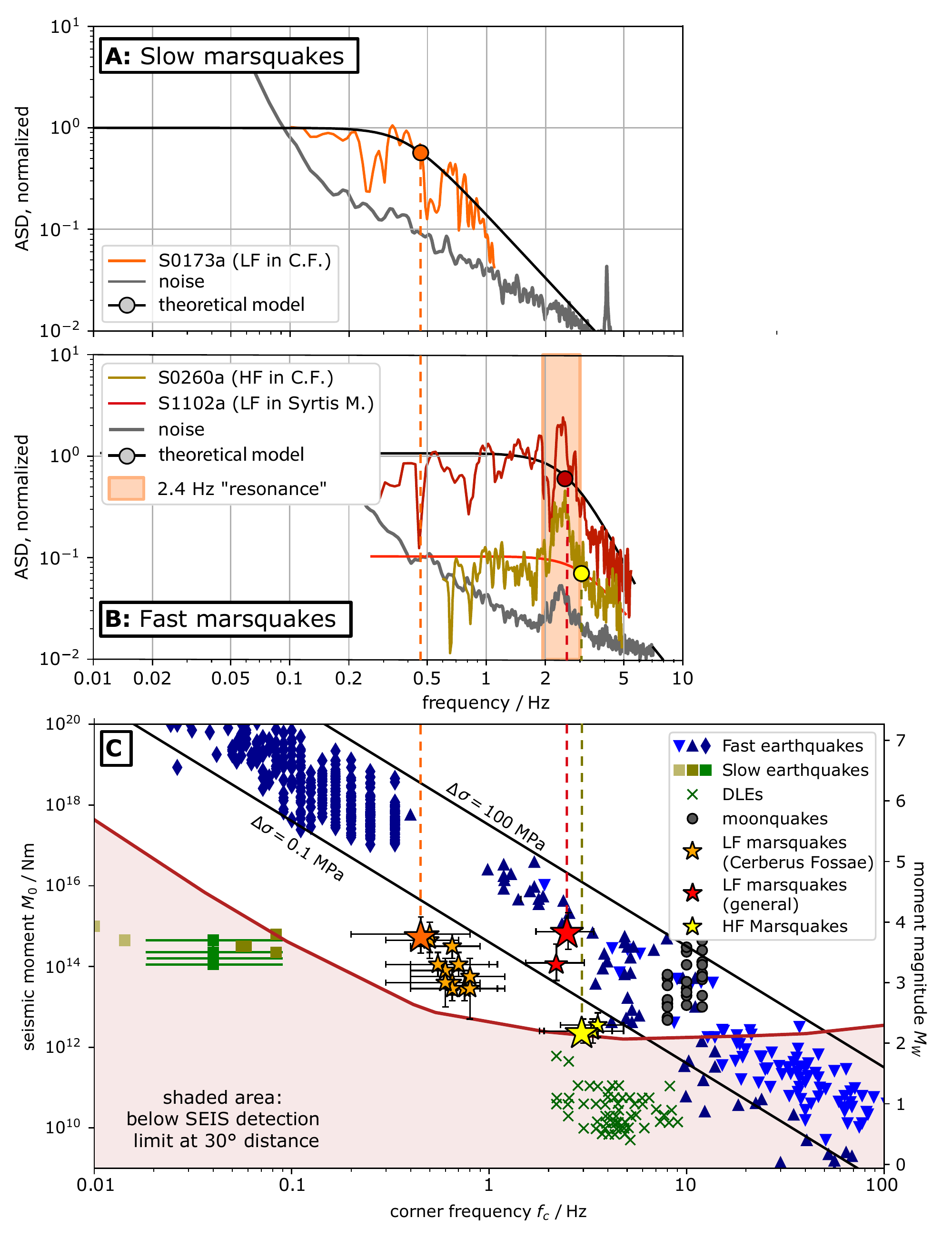}
\caption{Energy ratio between radial and transversal horizontal component for HF envelope stack (a) before and (b) after re-alignment. The energy maximum at a backazimuth of $78 \pm 12^{\circ}$ (dashed lines) corresponds to P-wave energy from the central part of Cerberus Fossae. The time axis is relative to an arbitrary offset used before alignment, the Pg-arrival is thus marked.}
\label{fig:HF_direction}
\end{figure}

Figure \ref{fig:overview_map} shows the combined probability density of all LF quakes locations including distance and direction, and it peaks at Cerberus Fossae. The uncertainty on back-azimuth results in a relatively large geographical spread in lateral direction. But given that the distance spread of the observed marsquake cluster is small, it can a priori be assumed that their lateral spread is of similar magnitude, allowing to place all of these events into Cerberus Fossae.
The Cerberus Fossae grabens are located in two main regions \citep{perrin_geometry_2022}. Graben systems 1, 2, 3 are in a distance range of 18$^{\circ}$ -- 27$^{\circ}$ from InSight, while Graben 4 spans the range of 33$^{\circ}$ -- 39$^{\circ}$. Graben 5 is a more faint and less mature segment that bridges the two main regions. The fossa strike is 15$^{\circ}$ off the direction of the lander. Distance differences of LF event clusters can most easily be explained by different locations along the fossae. To identify such locations, we evaluate if the event cluster is consistent with certain locations along the fossae (Figure \ref{fig:CF_map}a,b) when varying the velocity-density structure models. In the "near" end-member model, the cluster of events would occur at the eastern end of Graben 1 or in Graben 5. In the "far" case, it would be placed on Grabens 5 and 4. In any case, the dominant location of marsquakes at the eastern end of the western section is compatible with the observation of east-ward decreasing maturity by \citep{perrin_geometry_2022}. 
A single event, S0325a \citep{clinton_marsquake_2021} would  be located at the eastern end of Graben 4 (not shown in Figure 1), but given that this event is outside the clusters and has a poorly determined back azimuth \citep{zenhausern_lowfrequency_2022}, it cannot be clearly attributed to Cerberus Fossae. We conclude that the majority of \textit{all} LF events on the InSight hemisphere cluster in central Cerberus Fossae (Figure \ref{fig:overview_map}). 

Localising HF events has not been attempted due to the strong scattering and the lack of ballistic arrivals so far. We integrate our directional observation with MQS distances to investigate if HF events could also originate in Cerberus Fossae. We map their distribution and while their distribution is broader c.f. LF events, we find that it indeed matches the extent of Cerberus Fossae (Figure \ref{fig:overview_map} and \ref{fig:CF_map}a). Thus, shallow HF seismicity appears to be spread over a large part of the Fossae, as opposed to the more focused distribution of deeper LF events.

\section*{Seismic Moment}
\subsection*{Estimating Seismic Moment Release}
MQS routinely estimates magnitudes \citep{bose_magnitude_2018,bose_magnitude_2021} and their uncertainties. For LF events, the magnitude uncertainty takes into account the distance uncertainty as well as the error in estimating the long-period amplitude \citep{bose_magnitude_2021}. As described in \citep{giardini_seismicity_2020, clinton_marsquake_2021}, identification of marsquakes is limited to times of low wind at InSight, which make up approximately one third of the total duration averaged over the mission. While detection probability is distance- and magnitude-dependent, the distance-dependency can be neglected here, because all events are within a small distance range. When estimating the total seismic activity rate of a region, one needs to take into account not only the detection probability, but also the inherent randomness of number of events per year \citep{knapmeyer_estimation_2019}. Following the procedure described therein, we estimate the total annual moment rate in Cerberus Fossae as 1.4-5.6$\times 10^{15}$ Nm/yr. Regarding Mars' seismicity as a whole, SEIS observed a total of 39 LF marsquakes on the InSight hemisphere rated as quality A,B or C  (A-C is highest to lowest) by MQS up to December 31, 2021 \citep{insight_marsquake_service_mars_2021}. Five of them are unequivocally located outside Cerberus Fossae, compared to 14 in Cerberus Fossae. The remaining 20 are cannot be located due to noise, and several of them could thus originate in Cerberus Fossae as well. From the observations so far, the small region of central Cerberus Fossae accounts for at least half of the seismic moment release of the whole InSight hemisphere.

\subsection*{Seismic vs Geological Deformation}
Following the morphological estimate of \citep{taylor_estimates_2013}, the formation of the Cerberus Fossae segments 1-4 requires deformation equivalent to a seismic moment, $M_{0, \textrm{total}} =2.1\pm 0.5 \times 10^{24}$ Nm or $\dot{M}_{0} =0.5-2.2\times 10^{17}$ Nm/yr since initiation of the spreading 5-20 Ma ago. This prediction exceeds our seismic observation by a factor of 50 and is a first indication that the current seismicity rate is not representative for the entire formation process. 

Next, we focus on the observed seismicity cluster which spreads over 400 km distance and a range of $\sim$20 km in depth (thus providing an area $A$). Using a shear modulus of $\mu=24$~GPa \citep{knapmeyer-endrun_thickness_2021} and assuming that all of the seismicity was extensional, the observation is equivalent to a slip rate of $\dot s_{\mathrm{seism}} = \frac{\dot M_0}{\mu A}=7-30 \times10^{-6}$~m/yr. \cite{vetterlein_structural_2010} find a geological deformation rate $\dot d_{\mathrm{geol}}= 5-73\times10^{-5}$~m/yr for Grabens 1 and 2 (Northern Cerberus Fossae in their article). For our central part, the CMU shows an age of 53-210~ka \citep{horvath_evidence_2021} and a throw of at least 100~m, equivalent to a slip rate of $\dot s_{\mathrm{geol}}=5-20\times10^{-4}$~m/yr over the last few 10~ka. We therefore conclude that the current seismic slip rate $\dot s_{\mathrm{seism}}$ explains only $1-10\%$ of the total deformation, $d_{\mathrm{geol}}$, preserved in the geological record.

The shallow seismicity associated with the HF events is at least a factor of 10 below the LF events due to their smaller magnitudes and it is distributed over a larger area. It can therefore not explain the discrepancy between geological deformation and observed seismicity.

\section*{A Case for Dike-induced Tectonic Activity in the Source Region}
\subsection*{Spectral Characteristics}
\begin{figure}
\centering
\includegraphics[width=0.75\linewidth]{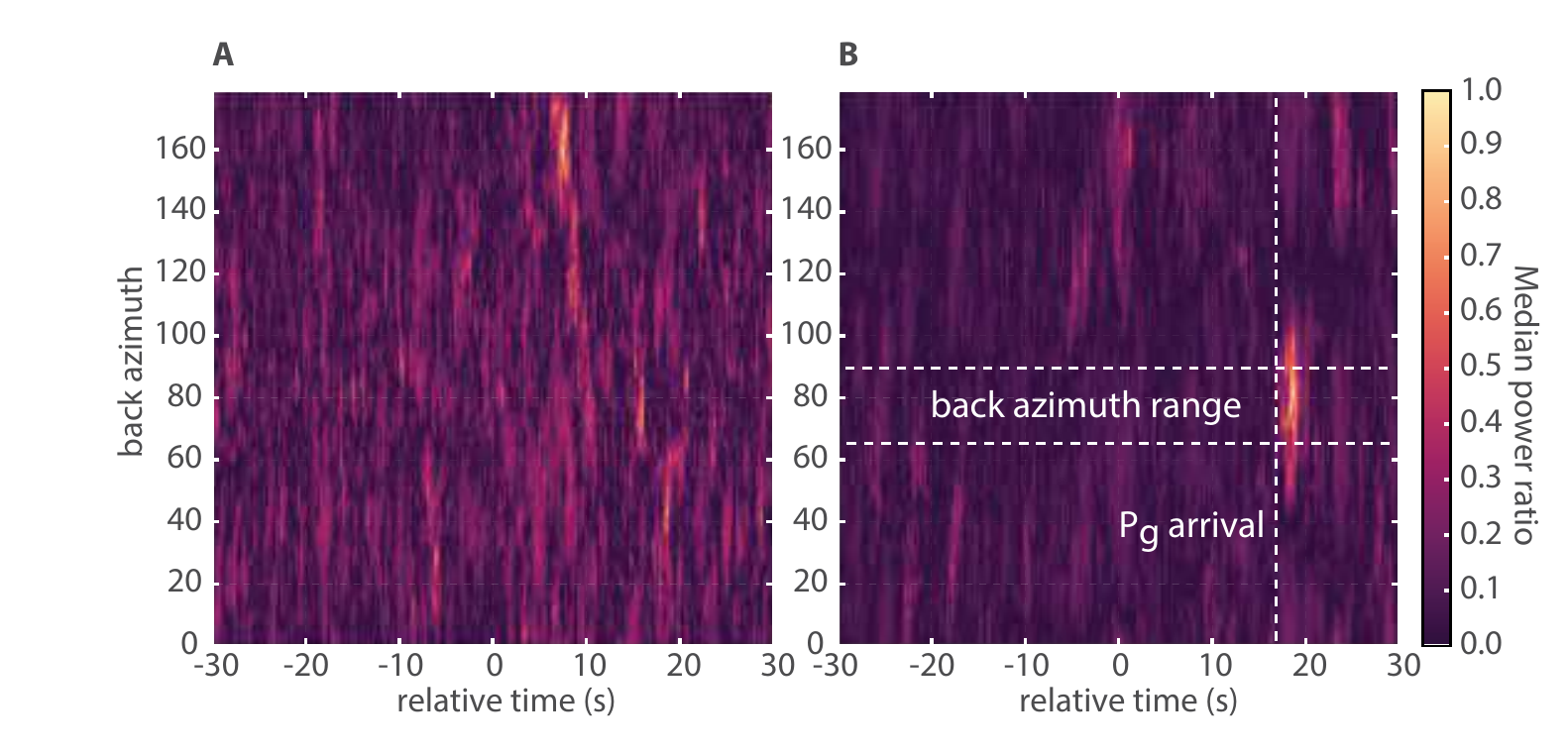}
\caption{Spectra of marsquakes and source parameters compared to terrestrial and lunar quakes. (A): Source spectra of Cerberus Fossae low frequency (LF) marsquakes S0173a. (B): Source spectrum for high frequency (HF) marsquake S0260a and distant LF marsquake S1000a. 
The source spectra were estimated from the vertical component for the P-wave window, normalized and corrected for attenuation, assuming $Q_{\mu}=1000$, to match P- and S-falloff. The black solid lines shows best fitting theoretical source spectra using a Brune model with exponent $n=2$ or $3$, circles are modelled corner frequencies. The background noise curves in (A) and (B) are from data before the P-wave arrival. The orange area in (B) is the local "2.4 Hz" subsurface resonance described in \citep{dahmen_resonances_2021-1, hobiger_shallow_2021}.
(C): Seismic moment $M_0$ vs corner frequency $f_c$ for different types of quakes observed on the Earth, the Moon and Mars. Blue symbols mark regular, "fast" earthquakes, following a cube law between seismic moment and corner frequency for 3 datasets of shallow earthquakes (in order of symbols: \citep{abercrombie_earthquake_1995, iio_scaling_1986, bilek_radiated_2004}). The brown squares mark a group of deep, slow events in Japan \citep{matsuzawa_source_2009}, the green crosses mark slow events related to volcanism, observed in Germany \citep{hensch_deep_2019}; gray dots are fast shallow moonquakes \citep{oberst_unusually_1987}.
Black lines are $f_c$ values for stress drops of 0.1 and 100 MPa for $\beta=3$~km/s.
HF marsquakes (yellow stars), as well as LF marsquakes outside Cerberus Fossae (red star) follow the $f_c\propto M^{-3}$ trend of earthquakes, while Cerberus Fossae LF events are significantly slower.}
\label{fig:scaling}
\end{figure}

The duration of a quake, whether on Earth or Mars, places a limit on the coherent high-frequency seismic energy radiated from it. This is typically expressed via the corner frequency in the source spectrum, $f_c$, above which displacement amplitude, $A(f)$, decreases as $f^n$, where $2<n<3$ \citep{brune_tectonic_1970}. Estimating the source spectrum of a quake is difficult from a single seismic record, because the high-frequency fall-off is affected by attenuation, both from intrinsic viscoelasticity, $Q_{\textrm{i}}$,  \citep{kane_quantifying_2011, trugman_source_2017} as well as scattering, $Q_{\textrm{scat}}$ \citep{aki_origin_1975}. For frequencies above 1 Hz, scattering has been found to affect P- and S-waves considerably, on Earth \citep{yoshimoto_frequency-dependent_1993} and on the Moon \citep{nakamura_seismic_1982}. Below 1 Hz, both attenuation mechanisms affect S-waves significantly stronger than P-waves due longer propagation time of S-waves. From a single seismic record, one can isolate source effects by correcting the observed spectra for different values of intrinsic shear wave attenuation $Q_{\textrm{i}}=Q_{\mu}$ until the P- and S-spectra match. Doing that, we find that the P- and S-wave spectra of LF events cannot be explained by effects of intrinsic attenuation alone, but show a strong source imprint. As an example, Figure \ref{fig:scaling}a shows the P-wave spectrum of the high-SNR event S0173a corrected for an average $Q_{\mu}=1000$, requiring a corner frequency $f_c = 0.45 \pm 0.15$~Hz. For all investigated LF events in Cerberus Fossae, we find that $0.45<f_c<0.95$~Hz. As shown in Figure \ref{fig:scaling}c, this is significantly less than the values found empirically for $M_{\textrm{W}}\approx3$ earthquakes \citep{abercrombie_earthquake_1995, allmann_global_2009}, which is 2-10 Hz. 
In comparison, LF marsquakes outside Cerberus Fossae, specifically a recent marsquake in the Tharsis region, S1000a (Fig \ref{fig:scaling}b), but also events at distances of 3000-4000 km (S0185a, S1102a, red stars in Fig \ref{fig:scaling}c) show significantly higher values of $f_c>1.5$~Hz. The corner frequency of the shallow HF events is significantly higher (see figure \ref{fig:scaling}b). For the largest HF events, a roll-off in displacement spectrum is observed above 3 Hz, which puts a lower limit on the $f_c$, given the unknown attenuation of the upper crust. While their magnitudes range from $1.5<M_{\textrm{W}}<2.5$, this $f_c$ is still at the low end of terrestrial quakes, although it must be noted that their magnitude is affected by less certain due to the complicated propagation mechanism. 

A feasible explanation for the observation of "slow" quakes is a significantly reduced shear wave velocity, $\beta$, in the source region, because in classic models $f_c$ scales linearly with $\beta$ \citep{brune_tectonic_1970}, i.e., 
\begin{equation}
    f_c \propto \beta \sqrt[3]{\Delta \sigma / M_0}.
    \label{eq:cornerfreq}
\end{equation}
where $\Delta \sigma$ is the stress drop. 
InSight observed low $\beta$ of 1.3-1.8 km/s in the uppermost ten km below the lander as derived using receiver functions \citep{lognonne_constraints_2020, knapmeyer-endrun_thickness_2021} and autocorrelations \citep{compaire_autocorrelation_2021}, which is a factor of $\sim$2 below the value in terrestrial crustal models \citep[e.g.][]{laske_update_2013}. If the hypocenters were located within this layer, this could explain a factor of 2 in corner frequency compared to the bulk of terrestrial crustal earthquakes, but hardly the observed 5-10. Also, so far all published results agree on LF event depths of at least 15 km \citep{brinkman_first_2021, stahler_seismic_2021,giardini_seismicity_2020}, where $\beta$ is comparable to terrestrial values.

Local weakening, e.g. due to warmer materials nearby a dike in the source region, could present a second effect leading to observations. The stress drop ($\Delta \sigma$ in eq. \ref{eq:cornerfreq}) describes the difference of the shear stress on the fault plane before and after the quake. It can be derived analytically for simple fault geometries in homogeneous media, and otherwise represents an empirical term \citep{abercrombie_resolution_2021}. As a general rule, low values for $f_c$ and thus $\Delta \sigma$ are found in volcanic settings, where material is heated and close to ductile behaviour \citep{giampiccolo_attenuation_2007}. 
This would require the hypocenters to be located in a zone of increased temperature possibly close to a magma chamber feeding shallower dikes (Figure \ref{fig:sketch}). This explanation is consistent with the depth estimates for LF and HF events. The deep LF events are closer to the heat sources and thus "slower" quakes than the more surficial and fast, brittle HF events. 
A third possible explanation would be that the lower gravity and temperature of the Martian crust will reduce yield strength and could thus lead to generally lower stress drops and therefore $f_c$. However, this general trend is not found on Earth, e.g. when comparing earthquakes at different depths \citep{vallee_source_2013} or tectonic settings \citep{abercrombie_resolution_2021}, and shallow moonquakes show even higher $f_c$ values  \citep{oberst_unusually_1987}. Finally, as stated above, large marsquakes outside of Cerberus Fossae have significantly higher corner frequencies, which fall within the range expected for terrestrial quakes of similar size (red line in Figure \ref{fig:scaling}b). This observation confirms that Cerberus Fossae events are different and, combined with the very frequent and localized observations of quakes from this source region, highlights the unique setting of Cerberus Fossae. 

\subsection*{Recent Volcanism in Cerberus Fossae}
The Cerberus Fossae Mantling Unit (CMU) near Zunil crater has been interpreted as a recent source of volcanism \cite{horvath_evidence_2021}, based on the symmetric distribution of dust and streaking of secondary craters away from the structure (Figure \ref{fig:CF_map}d). The dust and streaking directions overlay that of Zunil crater itself, dated at 0.1-1 Ma. Detailed age estimates of the CMU range between 53-210 ka, making it one of the youngest features mapped on the Martian surface to date \cite{horvath_evidence_2021}. Because there is no sign of younger volcanism locally, the structure must be considered dormant or inactive. 

Terrestrial volcanoes of this age can still be in an active state with ongoing fluid motion identified by seismic activity \cite{hensch_deep_2019}. The observation of marsquakes around the CMU is thus intriguing and we compare our observations with two categories of seismic activity near dormant volcanoes on Earth (for an overview, see e.g. \cite{chouet_multi-decadal_2013}): 
(1) Deep low frequency events (DLEs), which are "slow" quakes of magnitudes $<2$. These events typically occur in swarms, i.e. week- to year-long activity bursts. (2) Volcanic tremor, long-duration, monochromatic signals. 
In comparison, our events on Mars do not quite match any of these two categories. 
While martian LF events are abnormally slow, they do not qualify as DLEs, because they are too large in magnitude (which would require very significant subsurface magma motion \citep{kedar_analyzing_2021}). Their recurrence rate also shows no deviation from a stationary Poisson process, unlike DLEs, which occur in swarms.
We note that the seismometer distances of 1500 km or more lacks the resolution to observe very slow events, including DLEs below magnitude 2.5 (region below the red line in Figure \ref{fig:scaling}). We therefore cannot rule out signals from fluid motion, but discussed event observations are not consistent with Earth-like DLEs. \citep{kedar_analyzing_2021} investigated whether several LF marsquakes could be explained by volcanic tremor, but found that very large magma flow rates over short time windos would be needed. This is in apparent contradiction to the low number of marsquakes observed so far and the lack of surficial expressions of active volcanism. Also, the events discussed in our study show very clear P- and S-arrivals, unlike the more emergent signals of tremor on Earth.
In summary, we do not find evidence of tremor or generally fluid motion in the seismic data. However, we find an abnormally low $f_c$, indicative of slow rupture. 
\subsection*{Inferred Fracture Size}
Lastly, under the assumption of a circular source, $f_c$ allows to infer the source radius as $r=0.38\frac{\beta}{f_c}$ \citep{madariaga_dynamics_1976}. Within the range of $\beta$ between 2-3 km/s, we obtain rupture plane radii of 1200--1800~m. This is below the minimum size of mapped surface segments within the Cerberus Fossae (5-10 km \citep{perrin_geometry_2022}), suggesting that marsquake size is not primarily limited by fracture geometry and that over longer observation times, InSight could indeed observe significantly larger marsquakes.
The shallower HF quakes likely happen in the uppermost, low-$\beta$ layer \citep{knapmeyer-endrun_thickness_2021}, with source radii between 150 and 300~m. 

\section*{Discussion: The Evolution of Cerberus Fossae}
Seismic data confirms the ongoing opening of the Cerberus Fossae segments on Mars. Seismicity at 15-50~km depth with slow rupture processes suggests an extensional stress regime located in a warm source region. 
If the seismicity of the central event cluster is originating from the CMU, the observed seismic strain rate is far too low for a constant, slow opening of this feature \citep{horvath_evidence_2021}. This is in agreement with the hypothesis that it was created rapidly in a volcanic eruption 53-210 ka ago. 

The lateral distribution of events shows focused seismic activity in the center of Cerberus Fossae with generally low current seismicity compared to inferences from the geological record. 
This indicates that the opening rate of the grabens has not been constant, opposite to what was assumed by \cite{taylor_estimates_2013}. Instead, most segments likely opened rapidly and became mostly passive after a short time. The seismic observation period is short and our data merely represent a snapshot of the overall seismicity of Cerberus Fossae. Nevertheless, the fact that we do not see LF seismicity in most of Cerberus Fossae, specifically in the fractured western part where the largest deformation can be found, suggests a dynamic process that ceases after an initially active phase and continues to propagate eastward. 

\begin{figure}
\centering
\includegraphics[width=\linewidth]{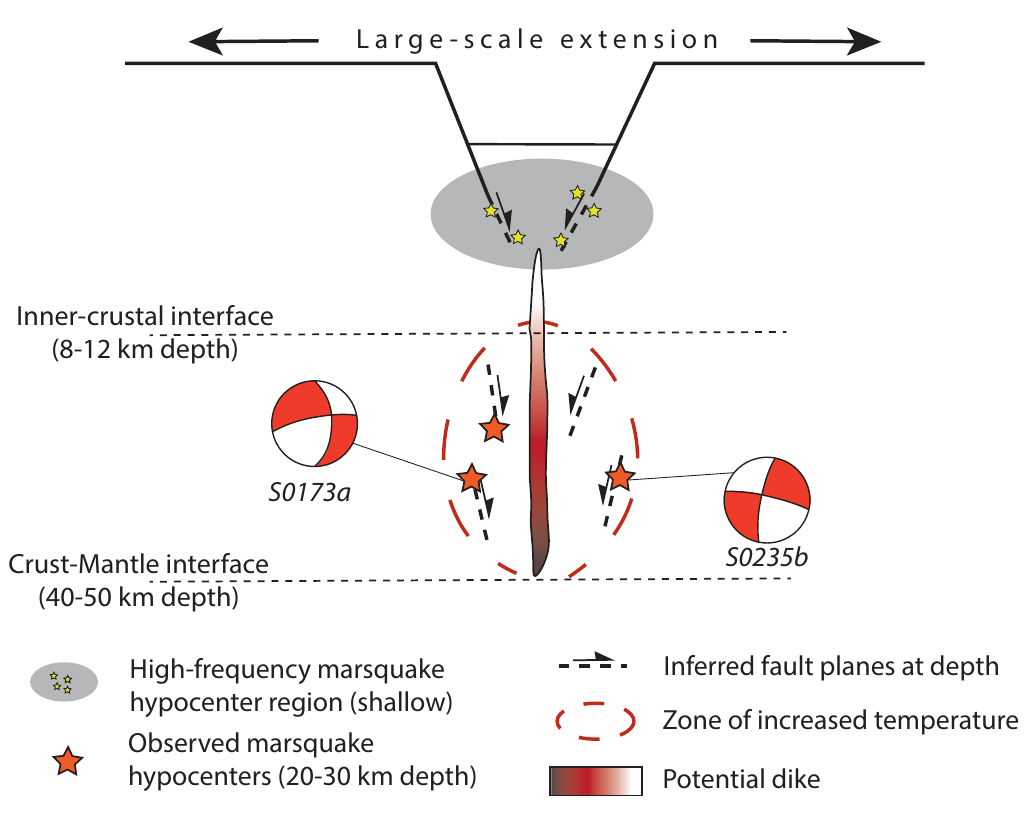}
\caption{Sketch of an active part of Cerberus Fossae viewed from East. The low frequency marsquake depths indicate faulting in the lower part of the crust, the low stress drop suggests that hypocenters (red stars) are located in the zone of increased temperature (red dashed) around a (recently) active dike at depth. Event S0235b has been previously located to the North of the fault, S0173a to the South. In combination with the focal mechanisms inferred in \citep{brinkman_first_2021}, rupture planes dipping towards the dike are plausible. In the shallow part, high frequency marsquakes are caused by residual stress on the flanks of the graben (yellow stars). }
\label{fig:sketch}
\end{figure}

We propose that the shallow seismicity from HF events is created by superficial ruptures due to the graben structure itself, possibly the subsurface continuation of the graben flanks (Figure \ref{fig:sketch}). The rapid rupture associated with these quakes is not consistent with sources such as landslides or other mass wasting processes. More likely it is caused by the release of residual stress. A modulation of the HF quake rate with a period of one Martian year was found to be consistent with the variation of solar illumination in equatorial latitudes \citep{knapmeyer_seasonal_2021}. Given that significant parts of Cerberus Fossae are deep enough to be shaded over half of a Martian year, this is at least a plausible correlation and consistent with shallow sources. However, a physical model connecting the two factors, illumination and quake rate, is still missing.

On a more global picture, the clearly localized seismicity indicates that global contraction and therefore lithospheric compression are not the primary driver of contemporary tectonics as it had been previously proposed for Mars \citep{phillips_expected_1991, knapmeyer_working_2006}. Cerberus Fossae alone releases 1.4-5.6$\times 10^{15}$ Nm/yr seismic moment, a factor of 2-8 more than the Moon globally \citep{knapmeyer_estimation_2019}, where shallow seismicity has been identified on compressional faults \citep{watters_shallow_2019}.
 
The slow character of the Cerberus Fossae events requires a warm source region. To be close to ductile rheology, a temperature of $1000\pm100$~K is required for basaltic compositions \citep{plesa_present-day_2018, bergman_intraplate_1986}. Assuming a quake depth of $40\pm10$~km, this results in a local crustal thermal gradient $\Delta T / \Delta z = 20 \pm 2$~K/km in Cerberus Fossae and a local heat flow of $36\pm10$~mW/m$^2$ (assuming thermal properties of basalt \citep{clauser_thermal_1995}), a factor of 1.7 above the global average of $21\pm7$~mW/m$^2$ recently found by \citep{khan_imaging_2021}.

The distribution and character of marsquakes show that loading from the weight of Tharsis cannot exclusively explain the origin of Cerberus Fossae. Instead, dike intrusions likely support the tectonic processes in its specific location, weakening the crust locally and allowing the grabens of Cerberus Fossae to open. Across the solar system, a pattern emerges, where tectonics of the larger terrestrial planets: Mars, Venus and the Earth, is dominated by volcanism \citep{byrne_comparison_2020} instead of purely passive cooling and shrinking, as it is found on the smaller Moon and Mercury.

\section*{Acknowledgements}
We acknowledge NASA, CNES, partner agencies and institutions (UKSA, SSO, DLR, JPL, IPGP-CNRS, ETHZ, IC, and MPS-MPG), and the operators of JPL, SISMOC, MSDS, IRIS-DMC, and PDS for providing SEED SEIS data.  S.C.S. acknowledges funding from ETH research grant ETH-10 17-3. S.C.S., G.Z., and D.G. acknowledge support from ETHZ through the ETH+ funding scheme (ETH+2 19-1: “Planet MARS”). A.M. acknowledges support from ETH 19-2 FEL-34 and the Harvard Daly Postdoctoral Fellowship. C.P. acknowledges support from CNES as well as Agence Nationale de la Recherche (ANR-14-CE36-0012-02 and ANR-19-CE31-0008-08). W.B.B. was supported by the NASA InSight mission and funds from the Jet Propulsion Laboratory, California Institute of Technology, under a contract with the National Aeronautics and Space Administration (80NM0018D0004).
This is InSight contribution 233.

\section{Tables}
\clearpage

\begin{table}
\centering

\begin{tabular}{lllllllll}
\toprule
\multirow{2}{*}{Event} & \multirow{2}{*}{Quality} & \multirow{2}{*}{Distance [deg]} & \multirow{2}{*}{$M_{\textrm{W}}$} & \multirow{2}{*}{$f_c$} & \multicolumn{2}{c}{Back Azimuth [deg]}\\ 
\cmidrule(lr){6-9}
       &      &   &  &   & MQS & Uncertainty & Pol.-based  & Uncertainty    \\
\midrule
\multicolumn{9}{c}{Events in Cerberus Fossae}\\
S0173a   & A & 30.0     & $3.7\pm0.3$ & $0.45 \pm 0.15 $ & 91 & 79-102 & 88      & 78-103    \\
S0235b   & A & 28.7     & $3.7\pm0.2$ & $0.45 \pm 0.15$ & 74 & 66-88 & 77      & 64-100    \\
S0802a   & B & 30.0     & $2.9\pm0.2$ & $0.75 \pm 0.25$ & -  & - & 82      & 65-96   \\
S0809a   & A & 29.8     & $3.3\pm0.2$ & $0.7 \pm 0.3$ & 87 & 67-105 & 91      & 82-100    \\
S0820a   & A & 30.2     & $3.3\pm0.2$ & $0.55 \pm 0.25$ & 88 & 76-107 & 106      & 85-120    \\
S0864a   & A & 28.7     & $3.1\pm0.2$ & $0.6 \pm 0.2$ & 97  & 83-116 & 90      & 66-110   \\
S0916d   & B & 29.3     & $2.9\pm0.2$ & $0.95 \pm 0.35$ & -  & - & 97      & 41-114    \\
\midrule
\multicolumn{9}{c}{Events likely in Cerberus Fossae}\\
\textit{S0105a}   & C & 32.5      & $3.0\pm0.4$ & $0.5\pm0.2$  & -  & - & 112     & 95-133   \\
\textit{S0325a}   & B & 39.7      & $3.7\pm0.3$ & $0.5\pm0.2$ & -  & - & 57      & 43-73    \\
\textit{S0407a}   & B & 29.3      & $2.9\pm0.3$ & $0.7\pm0.2$ & -  & - & 57      & 43-169    \\
\textit{S0409d}   & B & 31.1      & $3.2\pm0.3$ & $0.5\pm0.2$ & -  & - & 70      & 50-90     \\
\textit{S0474a}   & C & 29.1      & $2.9\pm0.3$ & $0.6\pm0.2$ & -  & - & 97      & 72-123    \\
\textit{S0484b}   & B & 31.8      & $2.9\pm0.2$ & $0.6\pm0.3$ & -  & - & 100      & 80-120    \\
\textit{S0784a}   & B & 34.5      & $3.3\pm0.2$ & $0.8\pm0.3$& -  & - & 115     & 92-136    \\ 
\midrule
\multicolumn{9}{c}{Other marsquakes}\\
S1102d   & B & 74      & $3.6\pm0.2$& $2.85\pm1.0$ & 25 & 11-37 & 22      & 354-55 \\
S0185a   & B & 59.8    & $3.1 \pm0.3$& $1.8\pm0.6$ & -  & - & - & - \\
\bottomrule
\end{tabular}
\caption{Table summarising marsquake parameters. The marsquake events, type (BB = broad-band, LF = low-frequency), distance, $M_{\textrm{W}}$, quality (highest to lowest for A, B, C) and MQS back-azimuths are taken from the MQS catalog \citep{insight_marsquake_service_mars_2021}. MQS uncertainties are described in \citep{clinton_marsquake_2021}. Magnitude $M_{\textrm{W}}$ based on \citep{bose_magnitude_2021}. Polarization back-azimuth values and uncertainties are described in \citep{zenhausern_lowfrequency_2022}. Events in \textit{italics} have less certain back-azimuth estimates (see \citep{zenhausern_lowfrequency_2022}).}
\label{tab:marsquakes}
\end{table}
\printbibliography
\appendix
\subsection{Global fault map}
\begin{figure}
    \centering
    \includegraphics[angle=90,origin=c,width=0.7\textwidth]{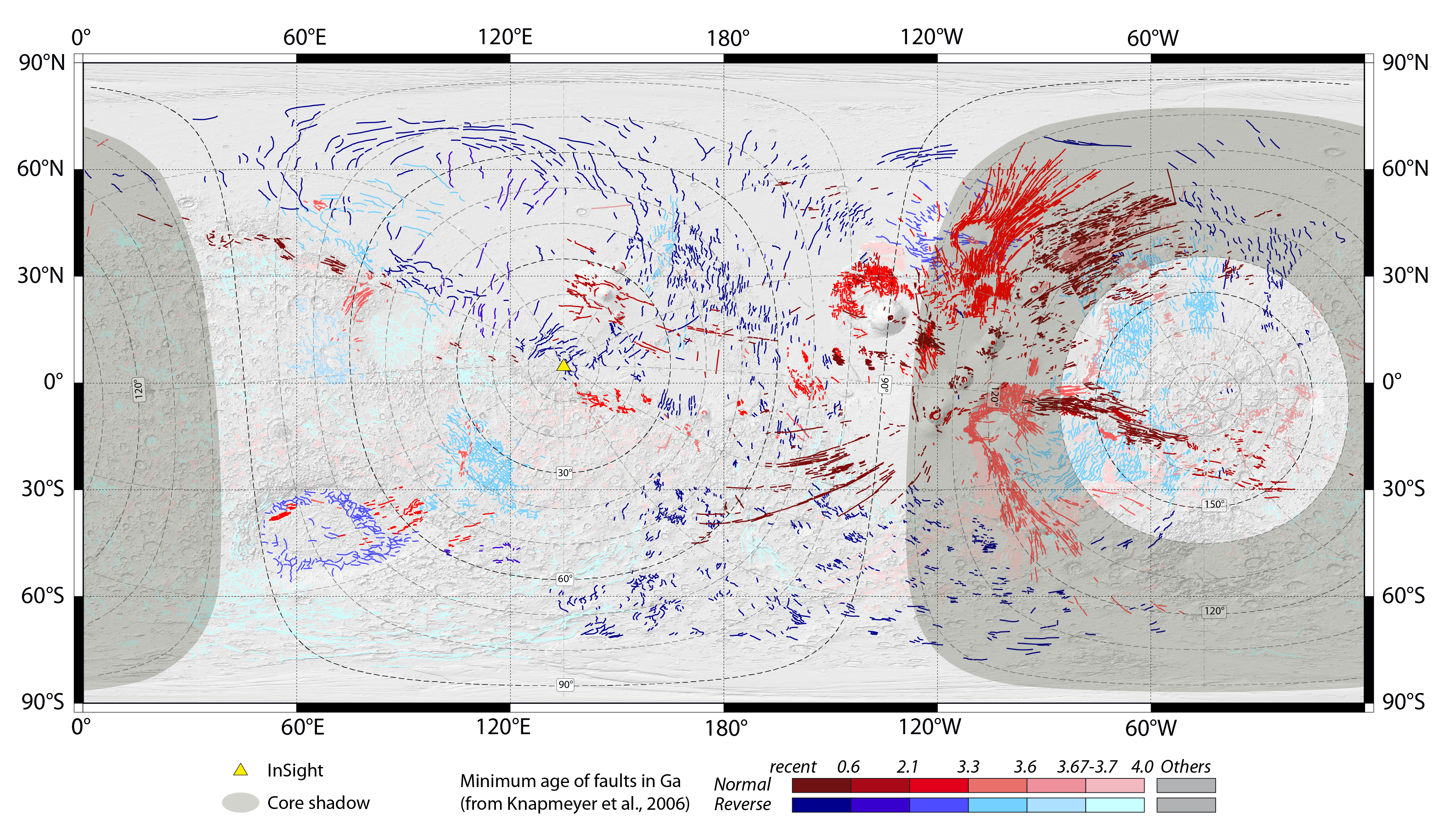}
    \caption{Global map of faults color-coded by minimum age  \citep{tanaka_geologic_2014, knapmeyer_working_2006, smith_mars_2001}. The darkened area marks the core shadow \citep{stahler_seismic_2021}, in which no direct body waves can be observed as seen from InSight. Thus event detection is significantly more difficult.}
    \label{fig:SM_global_map}
\end{figure}
\end{document}